\documentclass[a4paper]{elsart}
\usepackage[latin1]{inputenc}
\usepackage{subfigure}
\usepackage{graphicx}
\usepackage{epsfig}
\usepackage{wrapfig,rotating}
\usepackage{amssymb,amsmath,array}
\usepackage{epstopdf}

\begin{document}

\begin{frontmatter}
\title{Forward tracking at the next \boldmath{$e^+e^-$} collider\\ 
part I: the physics case} 
\author[valencia]{J. Fuster}
\author[cantabria]{S. Heinemeyer}
\author[valencia]{C. Lacasta}
\author[valencia]{C. Mari\~nas}
\author[cantabria]{A. Ruiz-Jimeno}
\author[valencia]{M. Vos\corauthref{cor}}
\corauth[cor]{Corresponding author. E-mail: Marcel.Vos@ific.uv.es}

\address[cantabria]{IFCA Santander, Avenida de los Castros s/n, E-39005 Santander, Spain }
\address[valencia]{IFIC (centre mixte U. Valencia/CSIC), Apd. Correos 22085, Valencia, Spain}

%


\begin{abstract}
In a series of notes we explore the detector requirements of the forward tracking region for a future $ e^+ e^- $ collider with a center-of-mass energy in the range from 500~GeV to 3~TeV. In this first part we investigate the relevance of the forward region for a range of physics processes that are likely to be relevant in such a machine. We find that many examples can be found where excellent performance of the forward detector system may lead to a considerable increase of the physics output of the experiment. A particularly clear physics case can be made for the reconstruction of electrons at small polar angle.
\end{abstract}
\end{frontmatter}

\section{Introduction}

The Standard Model (SM) of particle physics has a solid empirical basis. Much of the crucial experimental evidence was obtained in experiments studying the collisions of highly relativistic particles. Throughout the second half of the last century, ever larger devices were built to collide leptons (SPEAR, SLC, LEP) or hadrons (PS, SPS, Tevatron). With the start of data taking at the LHC (foreseen for this year) the center-of-mass energy accessible to man will reach well beyond a TeV. 

A consensus has emerged among particle physicists that an electron-positron collider with a luminosity of at least  $ 10^{33} cm^{-2}s^{-1}$ is the natural choice for the next collider to explore the energy frontier~\cite{consensus}. Two projects exist that pursue the creation of a linear electron-positron colliders (referred to as Linear Colliders or LC in the remainder of this note): 

\begin{itemize}
\item The proposal for an International Linear Collider (ILC~\cite{ilcrdr}) is based on existing super-conducting Radio-Frequency cavity technology. In the baseline design the ILC is envisaged to reach a center-of-mass energy of 500 GeV. The possibility to upgrade the ILC to a maximum of $ \sqrt{s} = $ 1~TeV has been a crucial requirement to the design since the very beginning. 
\item To reach even larger center-of-mass energies, the accelerating gradient obtained in the previous scheme is thought to be insufficient. The Compact Linear Collider (CLIC~\cite{clic,clicparams}) aims to open up the energy regime up to several TeV using a novel technology, where a drive beam is used to provide power to the RF cavities of the main Linac.
\end{itemize}

Several groups have performed exhaustive studies into the physics reach of the LC. A wealth of information can be found in the Technical Design Report for the TESLA project~\cite{tesla} and the physics workbook by the American Linear Collider Working Group~\cite{snowmass1}. The synergy with the LHC program is explored by the LHC/LC working group~\cite{weiglein}. The CLIC physics working group has analysed the reach of a 3~TeV $ e^+ e^- $ collider~\cite{clicphysics}.

The physics reach of experiments at the LC depends crucially on the performance of the detectors. A very active program exists aimed at the development of detectors for a linear collider~\cite{ilcdetectors}. Innovative approaches towards calorimeter, tracking and vertex detectors are pursued in detector R \& D collaborations. Three detector concept groups have prepared complete detector designs~\cite{ilcdetectorconcepts} for the ILC. In each of these the most important detector requirements are established on the basis of a careful maximization of the physics output.

In this note we will consider one of the most challenging areas of detector design for future linear $ e^+ e^- $ colliders: tracking in the polar angle range from approximately 5$^\circ $ - 30$^\circ $ (or, equivalently, 87 $ < \theta < $ 520 $ \rm ~mrad $, or 0.87  $ < \cos \theta < $ 0.996).

We want to maintain the focus of this note on the specific challenges of tracking in the forward region. We therefore choose, somewhat arbitrarity, to concentrate on the region with $ \theta < $ 30$^\circ $. This allows us to decouple, to a certain extent, from the discussion of the instrumentation of the central tracking. In all detector concepts, including those relying on a gaseous central tracking system, forward tracking relies on silicon-based devices arranged on a layout that consists of multiple disks.  

Neither will we discuss the challenge of instrumenting the very forward region ($ \theta < $ 5$^\circ $). Low-energy electron-positron pairs are produced very abundantly in the collision of the sub-micron size beams of an LC. In the strong solenoidal magnetic field in the tracking volume, these pairs accumulate in an approximately conical region in the forward and backward detectors. Therefore, the density of charged particles in the polar angle region with $ \theta < $ 5$^\circ $ is such that precision tracking is impossible.  

Reconstruction of charged particles in the polar angle range from \mbox{ 5$^\circ $ $ < \theta < $ 30$ ^\circ$} faces a number of specific challenges. The forward tracker in many experiments has not performed as well as the central tracker in key aspects like the momentum resolution or the material budget. The reasons for this situation are to be found at many levels of the detector design, from the solenoidal magnetic field in the tracking volume to the limitations of certain detector technologies and layouts and the routing of the services in the detector. A detailed discussion is beyond the scope of this note, but this issue will be addressed in depth in a future publication~\cite{forwardnote2}. Suffice it to say that in most experiments the forward tracking performance was compromised in favour of the central tracker.


 To strike the right balance between forward and central performance a detailed analysis of the impact on the physics reach of the experiment is required. In this note we will explore a number of physics cases where the forward region is particularly important. Thus, this note forms the basis of a series of documents that aims to provide a complete overview of the requirements of the forward tracking system for experiments at future $ e^+ e^- $ colliders. 

In the following some general features are highlighted using the example of some (potentially) very important processes for future colliders.   The list of benchmark processes, while far from exhaustive, includes processes from the standard categories, from SM measurements through Higgs boson and Supersymmetry studies to the searches for more exotic extension of the SM. Finally, in section~\ref{sec:conclusions} the most important findings are summarized and an outlook on future extensions of this study is given.

\section{Taking \mbox{ $ e^+ e^- \rightarrow Z/ \gamma^* \rightarrow f \bar{f}$ }; from LEP to the ILC to CLIC}
\label{sec:z}

\begin{figure}[htpb]
\centering
\includegraphics[width=0.48\linewidth]{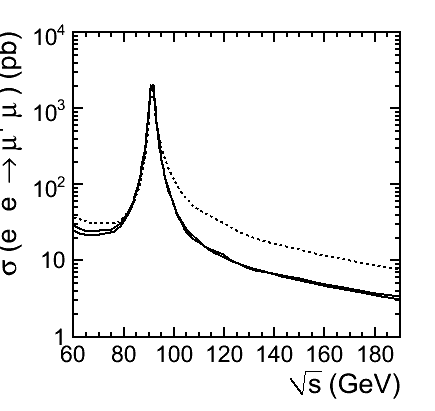} 
\includegraphics[width=0.48\linewidth]{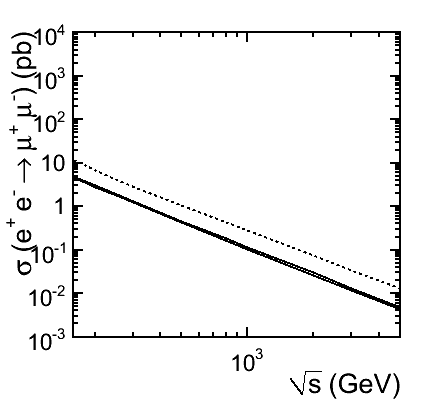}
\caption{Pythia~\cite{pythia} prediction for the production cross-section as a function of center-of-mass energy for the SM process $ e^+ e^-  \rightarrow Z / \gamma^* \rightarrow \mu^+ \mu^- $. The continuous line corresponds to events generated with ISR switched, while the dashed line includes ISR.} 
\label{fig:z}
\end{figure}

In this section we discuss the $ 2 \rightarrow 2 $ process $ e^+ e^- \rightarrow Z/ \gamma^+ \rightarrow f \bar{f}$, where $ f $ may denote all fermions. The cross section for the $ e^+ e^- \rightarrow \mu^+ \mu^- $ process is shown in figure~\ref{fig:z}. The most prominent feature is the Z-resonance: for collisions with a center-of-mass energy equal to the Z-mass the cross-section reaches several nb. The study of $ e^+ e^- $ collisions at the Z-pole formed the corner stone of the LEP-I program. 

In $ e^+ e^- $ collisions at larger and larger center-of-mass energy, the cross-section of the $ 2 \rightarrow 2 $ of figure~\ref{fig:z} shows a steady decline. In going from the maximum LEP-II energy (just over 200~GeV) to the multi-TeV range the SM cross-section drops over two orders of magnitude. As the center-of-mass energy increases the effect of real photons radiated off the incoming electron and positron legs becomes important. The cross section of the same process including initial state radiation, or ISR, is represented by the dashed line in figure~\ref{fig:z}.

\begin{figure}[htpb]
\centering
\subfigure[$ \sqrt s = $ 91~GeV]{\includegraphics[width=0.48\linewidth]{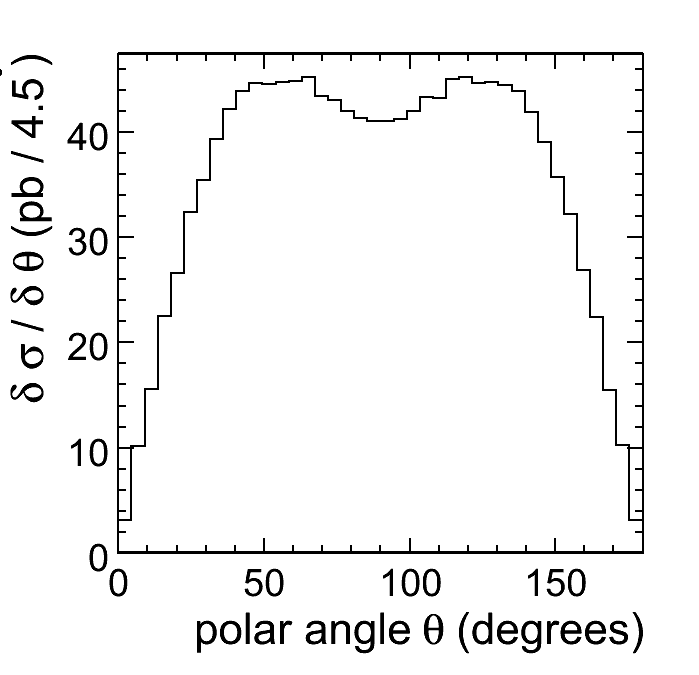}}
\subfigure[$ \sqrt s = $ 500~GeV]{\includegraphics[width=0.48\linewidth]{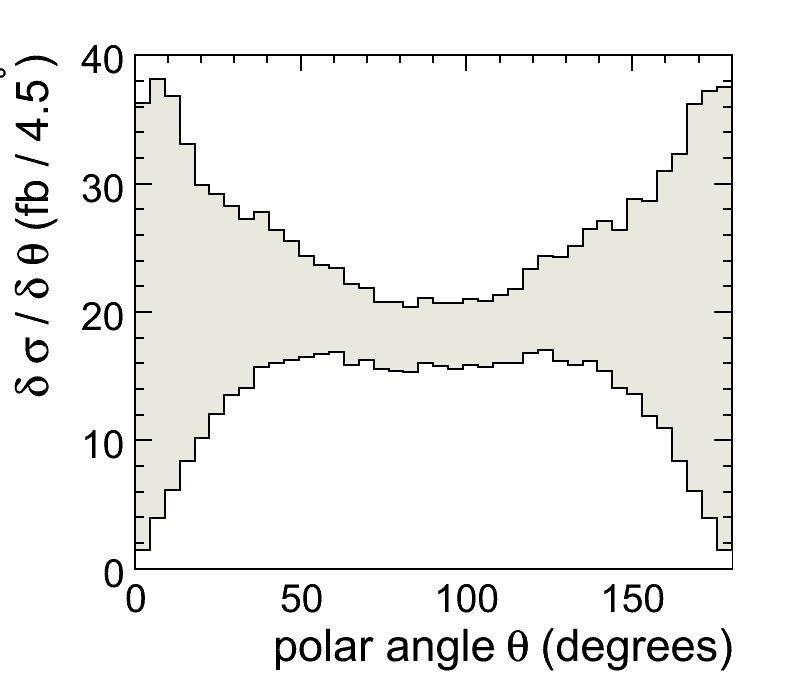}}
\subfigure[$ \sqrt s = $ 1~TeV]{\includegraphics[width=0.48\linewidth]{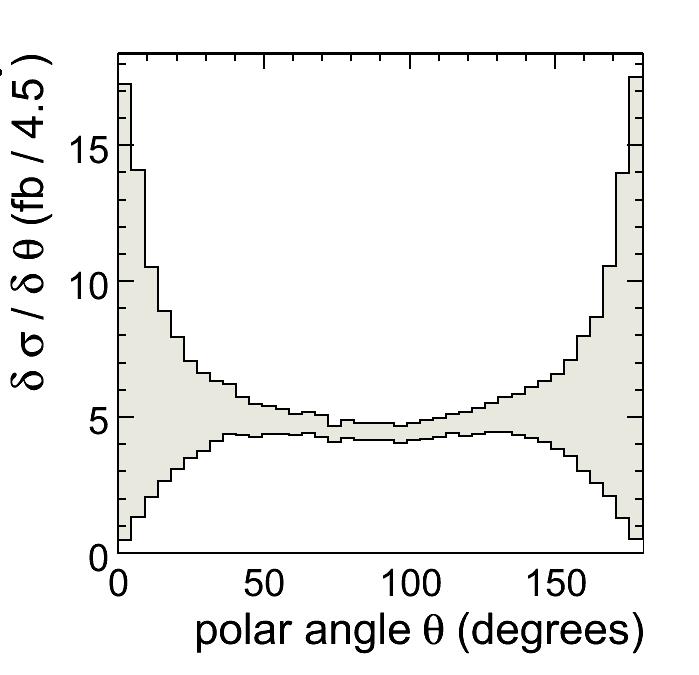}} 
\subfigure[$ \sqrt s = $ 3~TeV]{\includegraphics[width=0.48\linewidth]{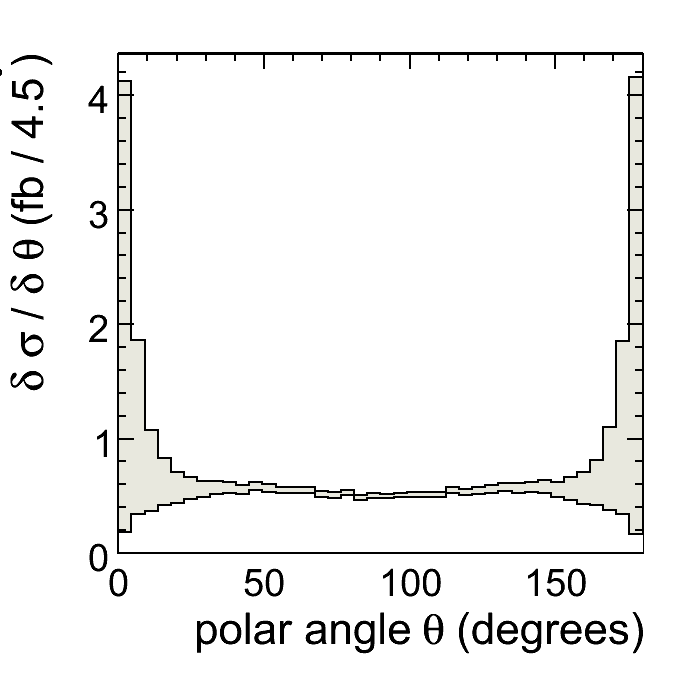}}
\caption{Pythia~\cite{pythia} prediction for the polar angle distribution of muons produced in the SM process $ e^+ e^-  \rightarrow Z / \gamma^* \rightarrow \mu^+ \mu^- $ for center-of-mass energies ranging from 91 GeV to 3~TeV. The grey fill area corresponds to the total cross section, including events with significant ISR, where the invariant mass of the $ \mu^+ \mu^- $ is less than two fifths of the beam energy. Note that the 91~GeV result is expressed in pb, while the other plots are in fb. }
\label{fig:z_polar_angle}
\end{figure}

The relevance of the forward region depends strongly on the center-of-mass energy considered. In figure~\ref{fig:z_polar_angle} the polar angle distribution of the final state muons is shown for four center-of-mass energies, from the Z-pole (the upper left panel) to 3~TeV (lower right panel). The histogram with the grey fill colour represents the total cross-section, including the effect of ISR. The histogram with the white fill area corresponds to events where the invariant mass of the outgoing muon pair is close to the nominal center-of-mass energy of the collider. 

The polar angle distribution for events where ISR is insignificant remains essentially unaltered between 500~GeV and 3~TeV. However, as the center-of-mass energy increases, the fraction of events with significant ISR increases. At the same time the polar angle distribution of these events becomes more and more peaked in the forward direction. Therefore, the forward detector region becomes more relevant as the center-of-mass energy is increased.

As a quantitative measure of the relevance of the forward region we will determine fraction $ P^X_{30} $ of final state particles of type X emitted in the forward direction (5$^\circ $ $< \theta < $ 30$ ^\circ $ or 175$^\circ $ $> \theta >$ 150$^\circ $). For the production of muon pairs at the Z-peak of figure~\ref{fig:z} a value of $ P^{\mu}_{30} = 18 $ \% is found. Less than 20 \% of the produced muons will be detected in the forward detectors. 


For larger center-of-mass energies the relevance of the forward region increases signficantly. For $ \sqrt{s} = $ 500~GeV the fraction of muons in the forward direction has nearly doubled: $ P^{\mu}_{30} = 33 $ \%. 

 \begin{figure}[htpb]
\centering
\includegraphics[width=\linewidth]{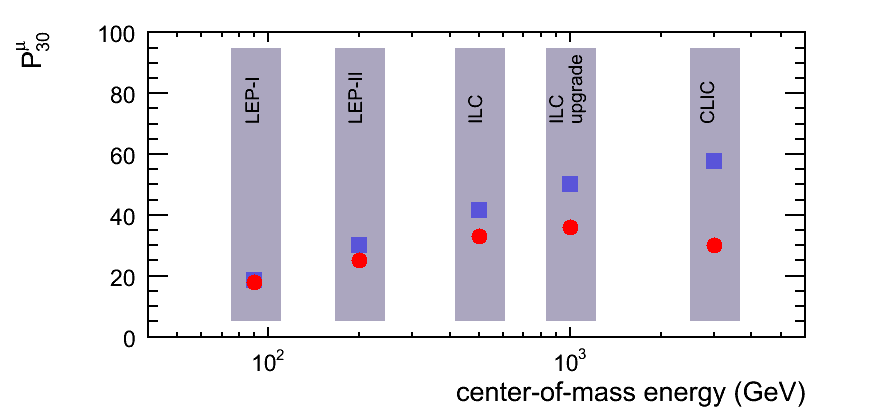}
\caption{Pythia~\cite{pythia} prediction for the fraction of muons emitted in the forward direction as a function of center-of-mass energy for the SM process $ e^+ e^-  \rightarrow Z / \gamma^* \rightarrow \mu^+ \mu^- $. The square markers indicate the fraction of muons with $ \theta < $ 30$^\circ$. The round markers were obtained by excluding the very forward region ($ \theta < $ 5$^\circ$).}
\label{fig:p30_vs_sqrts_z}
\end{figure}

In figure~\ref{fig:p30_vs_sqrts_z} the fraction of muons emitted in the forward direction is plotted as a function of center-of-mass energy. The trend towards greater relevance of the forward region continues when the center-of-mass energy is increased beyond 500~GeV. The peaks even move beyond $ \theta = $ 5$^\circ$ so that the forward fraction defined here does not increase further. For $ \sqrt{s} = 1 $~TeV, still within reach of an upgraded ILC, more than 50 \% of the muons is emitted at a polar angle of less than 30 $^\circ$, but only 36 \% is within the forward tracker. 

While the discussion is limited to muons, the polar angle distributions for other light fermions exhibit the same features. The numerical results for $ P^q_{30}$, where $ q $ may denote light, charm and bottom quarks, are within a few \% of those found for muons. The detection and identification of forward jets originating in charm or bottom quarks is particularly relevant for the measurement of the forward backward symmetry. Notable exceptions are the $ e^+ e^- $ final state that receives a significant contribution from the t-channel, and $ t \bar{t} $ that is discussed in the next section.  

We have seen that the final state products of the $ e^+ e^-  \rightarrow Z / \gamma^* \rightarrow \mu^+ \mu^- $ process are mostly emitted in the central detector when running at the Z-pole. The experiments  designed for LEP-I and SLC clearly reflected this preference: the design of SLD, OPAL, L3, ALEPH and DELPHI quite naturally focussed on the {\em central} detectors. For the next generation of experiments at center-of-mass energies of 500~GeV to 3~TeV, the relevance of the forward detector region increases significantly. Approximately one third of final state fermions (except top quarks) are emitted with a polar angle between 5$^\circ $ and 30$^\circ$.




\section{Multi-fermion final states}
\label{sec:multi}

As the center-of-mass energy of the collider increases, the relevance of $ 2 \rightarrow 4 $ fermion, $ 2 \rightarrow 6 $ fermion, and even higher multiplicity processes increases. The average number of jets in hadronic events, that was slightly less than 3 at the Z-pole, increases to nearly 5 for a 500~GeV ILC and 6.4 for 3~TeV collisions~\cite{clicphysics}. For high-multiplicity final states the final state fermions tend to become more isotropically distributed. As a result the probability for the full event to be contained in the central detector decreases. We will illustrate how this may lead to an increased relevance of the forward detector region using the example of SM $ t \bar{t} $ production.

Due to the large mass of the top quark the polar angle distribution deviates significantly from what was found for light quarks in section~\ref{sec:z}. At a center-of-mass energy of 500~GeV the fraction of top quarks emitted in the forward direction is quite small: $ P^t_{30} = 15 $ \%. For higher center-of-mass energies the forward fraction creeps up slowly, reaching 23 \% for a 3~TeV collider.

\begin{figure}[htpb]
\centering
\subfigure[any fermion]{\includegraphics[width=0.48\linewidth]{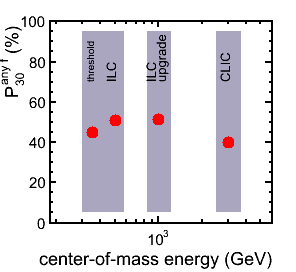}}
\subfigure[b-quark]{\includegraphics[width=0.48\linewidth]{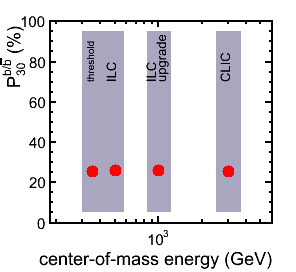}}
\caption{MadGraph~\cite{madgraph} prediction for the fraction of $ t \bar{t} $ events where one of its decay products is emitted in the forward direction. In the leftmost figure the fraction of events is represented in which at least one of the six fermions has $ \theta < $ 30 $^\circ $ or $ \theta <$ 150$^\circ $. In the rightmost figure only the b-quark and anti-quark are considered. The two points labelled threshold and ILC correspond to the $ t \bar{t} $ production threshold ($ 2 m_t = $350~GeV) and the nominal center-of-mass energy (500~GeV). }
\label{fig:p30_vs_sqrts_tt}
\end{figure}

In the Standard Model the top quarks decays with nearly 100 \% branching ratio into a bottom quark and a $ W^{\pm} $ boson. The latter decays into a quark and anti-quark or a charged lepton and a neutrino. Thus, the final state counts six fermions. For each of the decay products the probability to be emitted in the forward detector is similar to that of the top quarks. However, the polar angles of the decay products are only weakly  correlated. The fraction of events where at least one fermion is found in the forward region is shown in the leftmost panel of figure~\ref{fig:p30_vs_sqrts_tt}. For a 500~GeV linear collider $ P^{{\rm any}~f}_{30} = 51 $ \%: less than half of the events is fully contained in the central detector. Thus, precise reconstruction of $ t \bar{t} $ requires uniform jet energy resolution and lepton reconstruction performance over the full polar angle range. 

 The fraction of events where at least one of the b-jets is found in the forward region is shown in the rightmost panel of figure~\ref{fig:p30_vs_sqrts_tt}. The probability \mbox{$P^{b/\bar{b}}_{30} \approx 25 $ \%}, with a very slight dependence on the center-of-mass energy.

Future LC experiments, and in particular the detectors for CLIC, will have to cope with many-fermion final states. The example of $ t \bar{t} $ production illustrates that such events are rarely fully contained in the central detector. At the ILC, at least one of the six fermions is emitted in the forward region in over half of the events. In one of every four $ t \bar{t} $ events a b-jet is emitted with 5$^\circ $ $ < \theta < $ 30$ ^\circ$. Identification of very forward b-jet requires the vertex detector coverage to extend to very small angle.

%


\section{The t-channel}
\label{sec:susy}

So far, we have discussed {\em s-channel} processes. For $ e^+ e^- $ colliders in the center-of-mass energy range considered here, {\em t-channel} processes may well be very relevant. The polar angle distribution of final state objects in such processes exhibits pronounced peaks in the forward and backward direction. Therefore, quite generally, the study of {\em t-channel processes} requires excellent forward detector performance. In this section we provide a quantitive analysis of one benchmark reaction. We moreover study how the role of the forward detector evolves with increasing center-of-mass energy.  

\begin{figure}[!Hhtb]
\parbox{0.4\textwidth}{
\centering
\includegraphics[width=\linewidth]{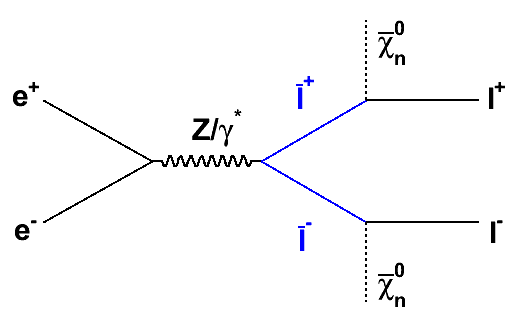}

}
\hfill
\parbox{0.4\textwidth}{
\centering
\includegraphics[width=\linewidth]{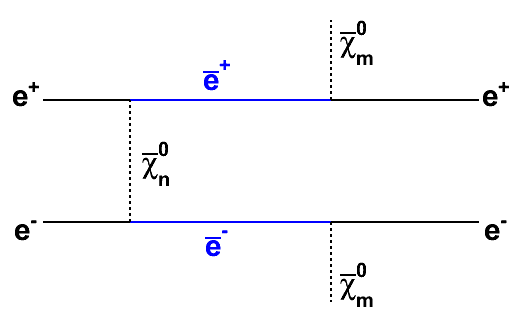}
}
\caption{The production processes for scalar leptons (sleptons) at an $ e^+ e^- $ collider.}
\label{fig:slepton_processes}
\end{figure}

Out of the possible {\em t-channel} reactions we choose to investigate scalar electron production, i.e. the rightmost diagram of figure~\ref{fig:slepton_processes}. This channel was studied in some detail by the SiD collaboration~\cite{demarteau_beijing}.

The rich phenomenology of supersymmetry in colliders at the TeV scale has been discussed extensively in the literature. If superpartners to the Standard Model exist and their mass are below $ \approx $2.5~TeV, the LHC is likely to reveal the existence of supersymmetry~\cite{weiglein}. The role of a future linear $ e^+ e^- $ collider in the 0.5-3~TeV range is to precisely determine the SUSY parameters. The LHC measurement of (charged) slepton (sleptons are the spin-0 partners to the SM leptons) and neutralino masses is expected to be improved by an order of magnitude at the ILC~\cite{weiglein}. 

Moreover, scalar lepton production can be of special relevance for the determination of the SUSY imposed relations on the $e-\tilde e-\tilde\chi^0$ and the $e^+-e^--\gamma$ coupling. Measuring the production and decay of scalar electrons precisely at an $e^+e^-$ collider is of particular importance. SUSY predicts the equality of the 
\begin{itemize}
\item[(i)]
electron-positron-photon coupling and the
electron-selectron-neutralino coupling (bino part),
\item[(ii)]
electron-positron-$Z$ coupling and the
electron-selectron-neutralino coupling (zino part).
\end{itemize}
This equality is only affected by higher-order corrections that grow
{\em logarithmically} with the {\em heavy SUSY mass scales}~\cite{sleptons1,sleptons2,sleptons3,sleptons4}.
Consequently, measuring the selectron couplings constitutes a deep 
test of SUSY as such. Furthermore, it could possibly reveal the scale of
otherwise unobservable large SUSY mass scales.

\bigskip

The MSSM possesses more than 100 free parameters. 
In order to study the features of the SUSY parameter space, simplifying
assumptions that may be more or less well motivated are often made, so
as to reduce the parameter space to a manageable dimensionality.
The most studied scenario is the Constrained MSSM (CMSSM), in which the soft
supersymmetry-breaking scalar and gaugino masses are each assumed to be
equal at some GUT input scale. In this case, the new independent MSSM
parameters are just four: the universal gaugino mass $m_{1/2}$,
the scalar mass $m_0$, the trilinear soft supersymmetry-breaking parameter
$A_0$, and the ratio $\tan{\beta}$ of Higgs vacuum expectation values.
Benchmark scenarios (mostly within the CMSSM) have been defined to
permit the comparison of various studies. The most common set are the 
Snowmass Points and Slopes~\cite{sps}, which we will also use in our study.

At an $ e^+ e^- $ collider, sleptons may be produced through the two processes indicated in figure~\ref{fig:slepton_processes}. The production of scalar muons $ \tilde{\mu}_R $ and $ \tilde{\tau} $-leptons at tree-level only involves the leftmost diagram $ e^+ e^- \rightarrow Z / \gamma^* \rightarrow \tilde{\mu}_R^+ \tilde{\mu}_R^- $ in figure~\ref{fig:slepton_processes}. 

\begin{figure}[htpb]
\centering
\includegraphics[width=1.0\linewidth]{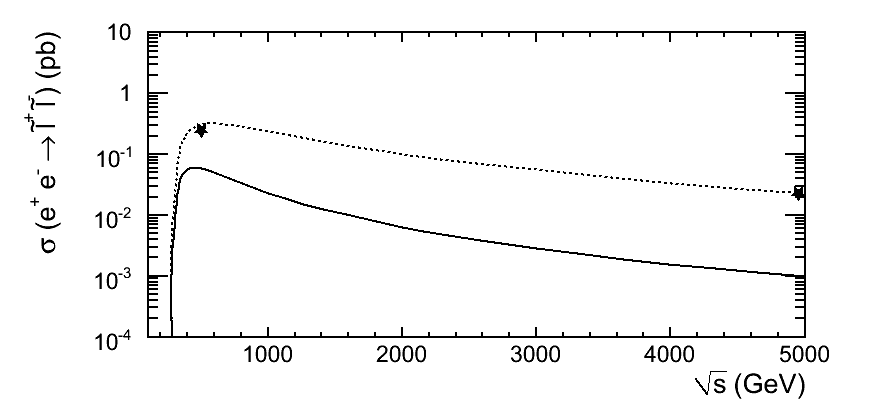}
\caption{MadGraph tree-level prediction for the production cross-section as a function of center-of-mass energy for scalar lepton production in SUSY benchmark point SPS1a~\cite{sps} $ e^+ e^- \rightarrow \tilde{\mu}_R^+ \tilde{\mu}_R^- $ (continuous line) and $ e^+ e^- \rightarrow \tilde{e}_R^+ \tilde{e}_R^- $ (dashed line). }
\label{fig:xsec_sleptons}
\end{figure}

The corresponding cross-section as a function of center-of-mass energy for scalar muons in benchmark point SPS1a is represented by the continuous line in figure~\ref{fig:xsec_sleptons}. In this benchmark point both the scalar electron and lightest neutralino are relatively light and slepton production is well within the reach of the ILC. Beyond a sharp threshold at $ \sqrt{s} = 2 m_{\tilde{\mu}_R}$ the production cross-section falls off rapidly. 

Just like the scalar muon pairs, scalar electrons are produced through an {\em s-channel} exchange of a $ Z$-boson or a virtual photon. However, in this case a number of additional diagrams like the rightmost diagram in figure~\ref{fig:slepton_processes} contribute. In the t-channel process a neutralino is exchanged. 

The dashed curve in figure~\ref{fig:xsec_sleptons} represents the production cross-section for the process $ e^+ e^- \rightarrow \tilde{e}_R^+ \tilde{e}_R^- $. Clearly, compared to scalar muons, the open {\em t-channel} diagrams lead to a much enhanced production of scalar electrons. The two points indicated by *'s indicate the cross-section obtained when the full$ e^+ e^- \rightarrow \tilde{e}_R^+ \tilde{e}_R^- \rightarrow e^+ e^- \tilde \chi^0 \tilde{\chi}^0$ is generated, limiting the polar angle of the electrons to $\theta > $ 5$ ^\circ$ and $ \theta <$ 175$ ^\circ $ and their transverse momentum to $ p_T > 10 $ GeV.

\begin{figure}[!Hhtb]
\parbox{0.48\textwidth}{
\centering
\includegraphics[width=\linewidth]{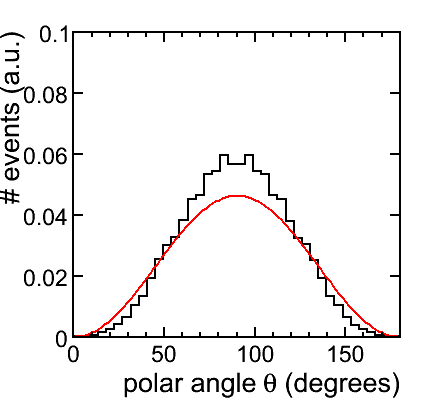}
}
\hfill
\parbox{0.48\textwidth}{
\centering
\includegraphics[width=\linewidth]{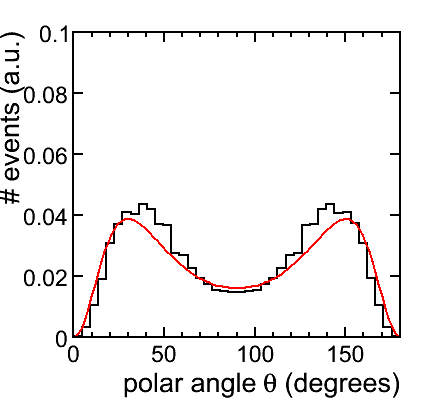}
}
\parbox{0.48\textwidth}{
\centering
\includegraphics[width=\linewidth]{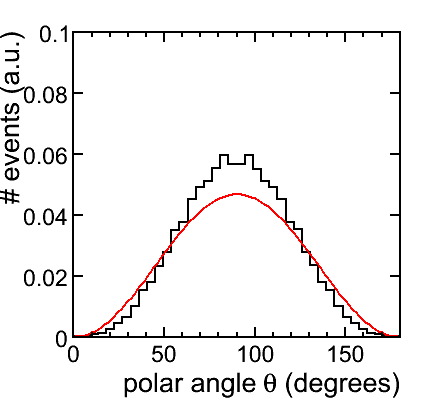}
}
\hfill
\parbox{0.48\textwidth}{
\centering
\includegraphics[width=\linewidth]{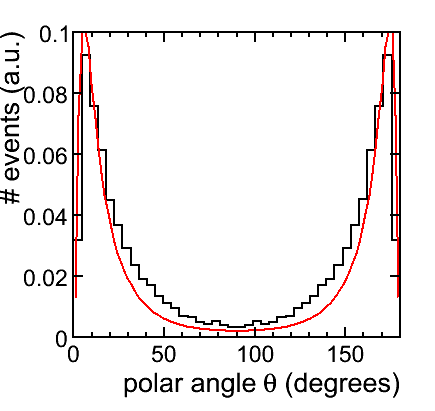}
}
\caption{The polar angle distribution for the $ e^+ e^- \rightarrow \tilde{\mu}_R^+ \tilde{\mu}_R^- $ (leftmost column) and $ e^+ e^- \rightarrow \tilde{e}_R^+ \tilde{e}_R^- $ process (rightmost column) for benchmark point SPS1a~\cite{sps}, where $ m_{\tilde{e}_R} = m_{\tilde{\mu}_R} = $ 144 GeV. The histogram represents the result from a MadGraph tree-level event generation. The continuous line corresponds to an analytical prediction of the  polar angle distribution (more details are given in the text). The uppermost and lowermost figures correspond to center-of-mass energies of 500~GeV and 3~TeV, respectively.}
\label{fig:susy_smuse_polar_angle}
\end{figure}


In figure~\ref{fig:susy_smuse_polar_angle} the polar angle distributions of the outgoing scalar leptons in benchmark point SPS1a are shown~\footnote{Note that scalar leptons are bosons: therefore the s-channel distribution is given by $ \sin^2 \theta $ rather than the $ 1 + \cos^2 \theta $ distribution for fermion pair production in section~\ref{sec:z}.}. The output of the MadGraph tree-level event generation (histograms) is compared to a analytical calculation of the polar angle distribution based on reference~\cite{tesiseduardo} (continuous line). This model uses the full Matrix Element of the s-channel diagrams. For scalar electron production the contribution from the four diagrams with neutralino exchange are taken into account.

In the leftmost column, the distribution for the pure {\em s-channel} process yielding a scalar muon pair is shown. The uppermost and lowermost plots correspond to center-of-mass energies of 500~GeV and 3~TeV, respectively. Independently of the collider properties, the s-channel process yields quite central distributions: the probability to find a scalar muon in the forward detector ($ \theta < $ 30$ ^\circ $, or 0.52 rad) is essentially negligible.  

The polar angle distribution of scalar electrons, in the rightmost panel, shows two very distinct peaks in the forward and backward direction. In this case the difference between a 500~GeV and a 3~TeV collider is quite significant. In a 500~GeV collider (uppermost figure) most of the scalar electrons are emitted at polar angles between 30$ ^\circ $ and 60$ ^\circ $. In a 3~TeV collider the peak is shifted to much smaller angles, and most of the scalar electrons are emitted at an angle below 30$ ^\circ $.

The analytical formulae are found to reproduce the qualitative features of the distribution correctly. Importantly, the $ \sqrt{s} $ dependence of the polar angle distributions of the t-channel process is reproduced quite accurately. The analytical result for fraction of scalar leptons emitted at small polar angle ($ \theta < $ 30$ ^\circ$ or $ \theta > $ 150$^\circ$) is found to be within 10 \% of the MadGraph result for all cases considered. 

\begin{table}
\caption{The fraction of scalar electrons emitted in the forward direction in nine SPS benchmark scenarios~\cite{sps}. Five center-of-mass energies ranging from 500~GeV to 3~TeV are considered. Open places in the table indicate cases where the scalar electron is not accessible.}
\begin{tabular}{|lcc|ccccc|}
\hline 
  \multicolumn{3}{|c|}{Benchmark point}  & \multicolumn{5}{c|}{$ P^{\tilde{e}_{R}}_{30} $  ( \% ) at LC with $ \sqrt{s}= $}  \\
  name & $ m ( \tilde{e}_R ) $ & $ m ( \tilde \chi^0_1 ) $          &   500~GeV & 800 GeV & 1~TeV & 2~TeV & 3~TeV \\ \hline
 SPS1a  & 135 & 99  & 30  &  46 & 54 & 70 & 73 \\ \hline 
 SPS2  & 1451 & 79  & -   &  -  & -  & -  & 10 \\ \hline 
 SPS3  & 178 & 160  & 20  &  38 & 48 & 63 & 70 \\ \hline 
 SPS4  &  416 & 118 & -   &  -  & 21 & 65 & 72 \\ \hline  
 SPS5  &  192 & 119 & 21  &  47 & 57 & 70 & 71 \\ \hline 
 SPS6  &  236 & 189 & 8   &  27 & 38 & 64 & 73 \\ \hline
 SPS7  &  127 & 161 & 25  &  35 & 43 & 65 & 73 \\ \hline 
 SPS8  &  176 & 137 & 24  &  44 & 47 & 66 & 72 \\ \hline
 SPS9  &  303 & 175 & -   &  26 & 42 & 61 & 67 \\ \hline
\end{tabular}
\label{tab:sleptons}
\end{table}

The fraction of scalar leptons emitted in the forward direction is listed in table~\ref{tab:sleptons}. Five different center-of-mass energies are considered for each of the SPS benchmark points. The result is found to depend strongly on the slepton mass, the masses of the neutralinos and the center-of-mass energy. The fraction of forward scalar electrons increases strongly with increasing center-of-mass energy. For a 500~GeV ILC less than a third of the scalar leptons are emitted in the forward direction for all accessible benchmark points. In a multi-TeV collider two thirds of scalar electrons will be produced with a polar angle 5$^\circ $ $ < \theta <$ 30$^\circ $. 

\begin{figure}[htpb]
\centering
\subfigure[SPS1a]{\includegraphics[width=0.48\linewidth]{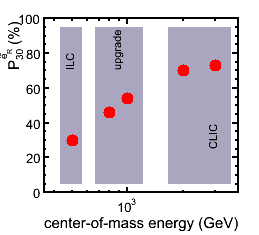}}
\subfigure[SPS9]{\includegraphics[width=0.48\linewidth]{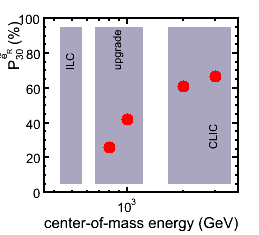}}
\caption{MadGraph~\cite{madgraph} prediction for the fraction of scalar electrons emitted in the forward direction for five different center-of-mass energies. Two panel correspond to two benchmark points: SPS1a and SPS9~\cite{sps}.}
\label{fig:p30_vs_sqrts_selectrons}
\end{figure}

In figure~\ref{fig:p30_vs_sqrts_selectrons} the dependence of $ P^{\tilde{e}_{R}}_{30} $ on the center-of-mass energy is depicted for two benchmark points.

It is instructive to see how two (more or less arbitrary) choices of a benchmark channel may lead to entirely different detector specifications. If scalar muon production is chosen as a benchmark channel a quite challenging requirement on the momentum resolution of central, high $p_T $ muons will be derived. A detector optimized for this analysis will therefore be equipped with a very precise central tracker. The performance of the forward tracker is largely irrelevant for this analysis. 

If instead we choose scalar electron production as our benchmark reaction the situation is radically different. A precise measurement of the scalar electron mass (through the reconstruction of the end-point) will require a very precise reconstruction of the electron momentum. Therefore, the forward detector region should be equipped with a precise, but extremely low-mass tracking system in combination with excellent electromagnetic calorimetry.

\section{Gauge boson pair production}
\label{sec:diboson}

\begin{figure}[!Hhtb]
\subfigure[]{\includegraphics[width=0.24\linewidth]{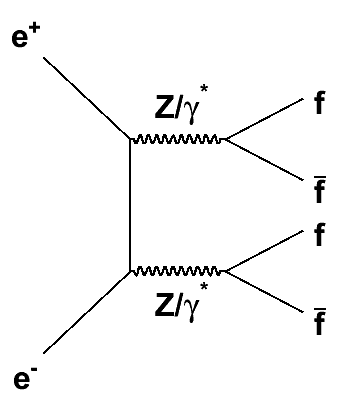}}
\subfigure[]{\includegraphics[width=0.24\linewidth]{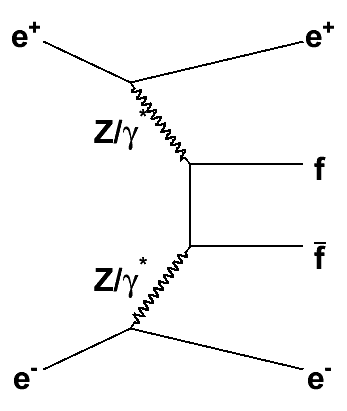}}
\subfigure[]{\includegraphics[width=0.24\linewidth]{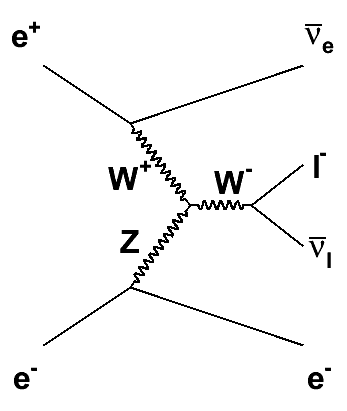}}
\subfigure[]{\includegraphics[width=0.24\linewidth]{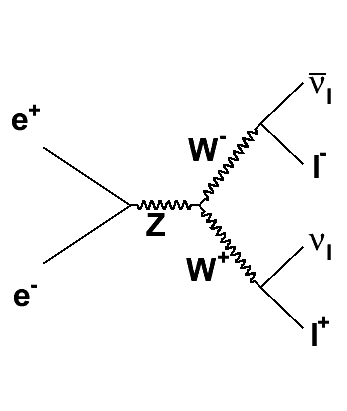}}
\caption{Some of the Feynman diagrams for the $ e^+ e^- \rightarrow f \bar{f} f \bar{f} $ process (a) and (b) and for the $ e^+ e^- \rightarrow l^+ l^- \nu_l \bar{\nu}_l$ process (c) and (d).}
\label{fig:zzwwdiagrams}
\end{figure}

 In this section we discuss the relevance of the forward detector region for two classes of $ 2 \rightarrow 4 $ processes. The very distinctive final state with four charged leptons is dominated by processes involving two Z-bosons like the leftmost diagrams in figure~\ref{fig:zzwwdiagrams}. The $ e^+ e^- \rightarrow W^+ W^- \rightarrow l^{+} l^{-} \nu_l \bar{\nu_l}$ process has a very considerable cross-section of several pb for center-of-mass energies in the range from 500~GeV to 1~TeV. For more details the reader is referred to Ref.~\cite{Denner}.

\begin{figure}[!Hhtb]
\subfigure[$ e^+ e^- \rightarrow l^+ l^- l^+ l^- $]{\includegraphics[width=0.48\linewidth]{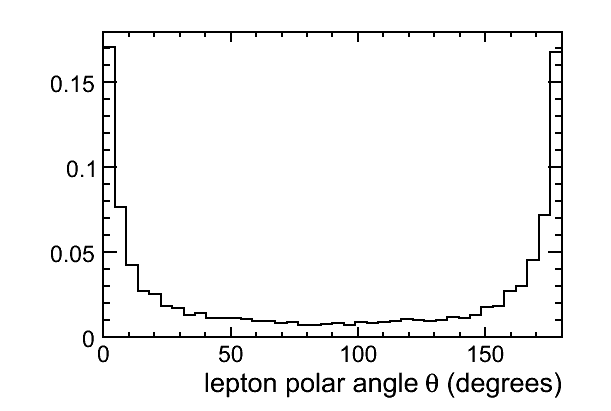}}
\subfigure[ $ e^+ e^- \rightarrow l^+ l^- \nu_l \bar{\nu}_l$]{\includegraphics[width=0.48\linewidth]{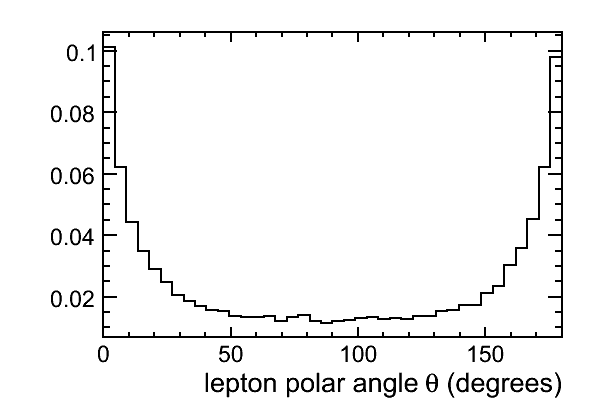}}
\caption{The polar angle distribution for the charged leptons in the final state of all tree-level diagrams yielding four charged leptons and two charged leptons and two neutrinos at a center-of-mass energy of 1 TeV. }
\label{fig:zzwwpolarangle}
\end{figure}

Standard Model processes yielding final states with two gauge bosons are an important source of background at an LC. The triple gauge boson vertex in this process (the rightmost diagrams of figure~\ref{fig:zzwwdiagrams}) could moreover be sensitive to physics beyond the Standard Model.

One should distinguish diagrams like (a) where the incoming electron and positron annihilate and the final state fermions can be chosen freely: $ l = e , \mu, \tau $ (and the same diagrams exist for neutrinos and quarks). In other diagrams, like (b), the flavour of two leptons in the final state is determined and the process is effectively $ e^+ e^- \rightarrow e^+ e^- f \bar{f} $. At larger center-of-mass energy the fraction of events with electrons and positrons (rather than muons or $ \tau $-leptons) increases from approximately 1/2 at 500~GeV to 2/3 at 3~TeV.

\begin{figure}[!Hhtb]
\centering
\subfigure[$ l^+ l^- \nu \bar{\nu} $ ]{\includegraphics[width=0.48\linewidth]{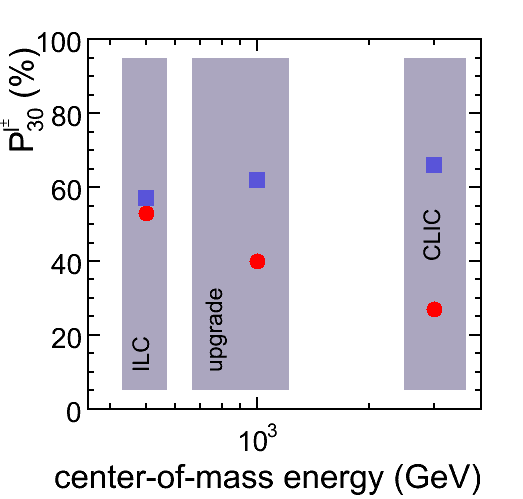}}
\subfigure[ $ l^+ l^- l^+ l^- $ ]{\includegraphics[width=0.48\linewidth]{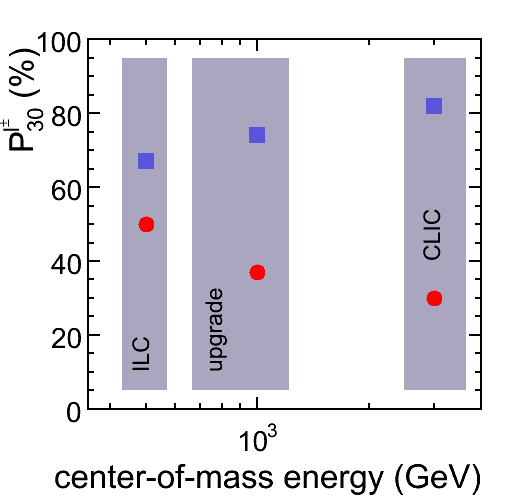}}
\caption{MadGraph~\cite{madgraph} prediction for the fraction of charged leptons emitted in the forward direction in $ l^+ l^- \nu \bar{\nu} $  and $ l^+ l^- l^+ l^- $ events. The round markers represent $ P^{l^{\pm}}_{30} $, while the squared markers correspond to the total fraction of forward charged leptons ($ \theta <$ 30$^\circ$).}
\label{fig:p30_vs_sqrts_ww_zz}
\end{figure}

The polar angle distribution of both processes is given in figure~\ref{fig:zzwwpolarangle} for a center-of-mass energy of 1~TeV. The events have been generated requiring a minimal $ p_T $ of 10~GeV and a minimum (maximum) polar angle of 1$^\circ$ (189$^\circ$). Both distributions exhibit strong peaks in the forward direction. At a center-of-mass energy of 500~GeV the probability for the electron or positron to be emitted in the forward direction is approximately two thirds. For other fermions this probability is 45 \%. At larger center-of-mass energy the tendency for electrons to be emitted in the very forward direction becomes very pronounced.

The fraction of leptons emitted in the forward region is presented in figure~\ref{fig:p30_vs_sqrts_ww_zz}. With increasing center-of-mass energy more and more electrons (and to a lesser extent also muons and $ \tau $-leptons) are emitted in the forward direction. The fraction of charged leptons with a polar angle $ \theta < $ 30$^\circ$ (square markers) increases steadily. Especially at large center-of-mass energy a very significant fraction of charged leptons escape detection in the forward tracker. The net result is that $ P^{l^{\pm}}_{30} $ decreases from over 50 \% to approximately 20 \% as the center-of-mass energy increases from 500~GeV to 3~TeV.

With this large probability for single leptons to be emitted in the forward direction, the fraction of $ WW $ and $ ZZ $ events that are fully contained in the central detector is negligible. Therefore, these processes form a very strong case for precise lepton reconstruction at low polar angle. The argument is particularly strong for electrons. It should be noted that despite their moderate transverse momentum ($ < p_T > \sim $ 50 GeV) these are quite energetic leptons: the average momentum is over 300 GeV for $ \sqrt{s} = $ 1~TeV.


\section{Higgs boson production}

\begin{figure}[!Hhtb]
\subfigure[Higgs-strahlung]{\includegraphics[width=0.3\linewidth]{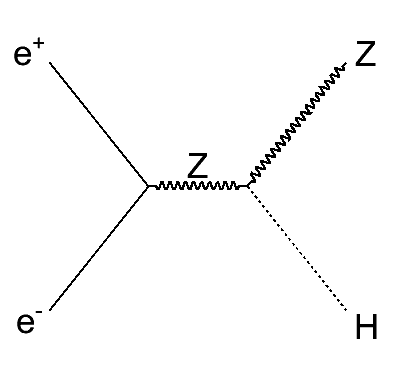}}
\subfigure[Z-boson fusion]{\includegraphics[width=0.3\linewidth]{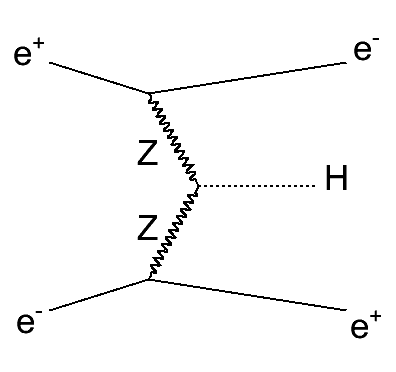}}
\subfigure[W-boson fusion]{\includegraphics[width=0.3\linewidth]{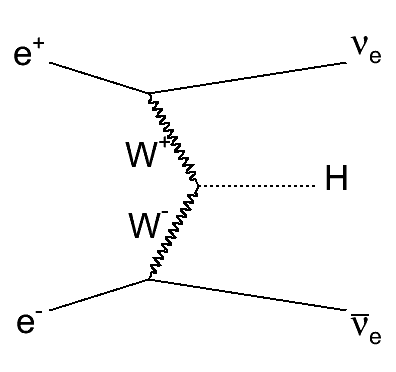}}
\caption{The production processes for the Higgs boson.}
\label{fig:higgs_processes}
\end{figure}

The Higgs-strahlung process of figure~\ref{fig:higgs_processes} (a) for a light SM Higgs boson is one of the most studied benchmark channels for an $ e^+ e^- $ collider operated at a center-of-mass energy of several 100~GeV. In the {\em recoil} analysis the Higgs boson mass can be measured independent of the decay of the Higgs boson itself. The mass is measured by reconstructing the decay of the Z-boson into $ \mu^+ \mu^-$. This analysis is one of the driving benchmarks for the design of the tracker. For a center-of-mass energy slightly above the sum of the Z-boson and Higgs-boson masses, the polar angle distribution of the muons from Z-decay is quite central ( $ P^{\mu}_{30} =$ 14 \% for $ \sqrt{s} = $ 250~GeV). Events with muons emitted in the very forward direction are typically removed from the analysis~\cite{Lohmann,li}. Therefore, little guidance for the design of the forward tracker is to be obtained from a study of the Higgs-strahlung process. 

 Besides the Higgs-strahlung process shown in the left diagram of
  figure~\ref{fig:higgs_processes}, Higgs production at the LC has strong contributions from two other processes. The final state in the central diagram is $H e^+ e^-$. Finally, the W-boson fusion process in the rightmost diagram leads to $ H \nu_e \bar{\nu}_e $.

\begin{figure}[!Hhtb]
\centering
\includegraphics[width=0.8\linewidth]{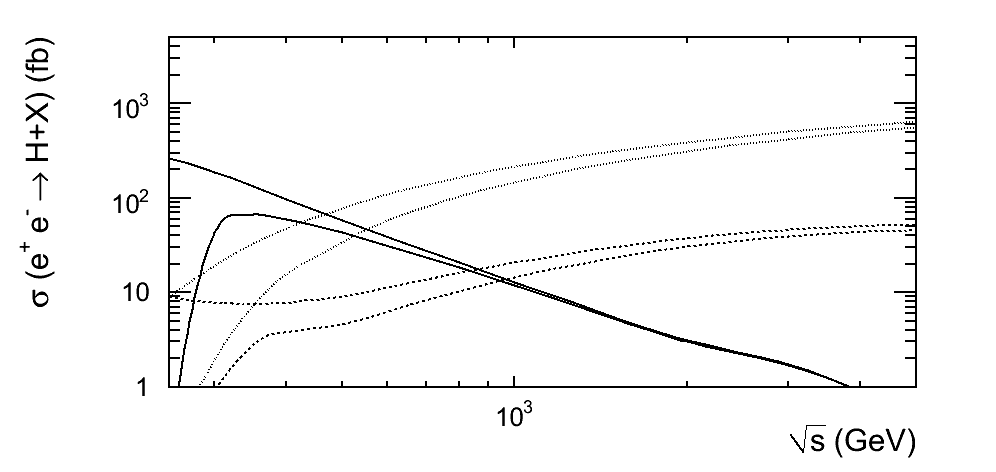}
\caption{The Higgs bosons production cross-section as a function of center-of-mass energy. The continuous curve corresponds to the Higgs-strahlung process, the dashed curve to ZZ-fusion and the dotted curve to WW-fusion. For each process the upper and lower curve represent Higgs boson masses of 120~GeV and 200~GeV, respectively. }
\label{fig:higgs_xsec}
\end{figure}

 The cross-sections of the three Higgs production processes are indicated in figure~\ref{fig:higgs_xsec}. Each pair of curves represents two Higgs masses: 120~GeV (the upper curve) and 200~GeV (the lower curve). The continuous lines correspond to the Higgs-strahlung process. The dashed and dotted lines correspond to the Z-boson fusion and W-boson fusion processes, respectively. While the cross-section for the Higgs-strahlung process drops rapidly (approximately as $1/s$), the cross-section of the vector boson fusion processes rises steadily (approximately as $ \sigma \propto \log{s} $) with increasing center-of-mass energy. For center-of-mass energies beyond 500~GeV the vector-boson fusion processes dominate the Higgs boson production cross-section.

\begin{figure}[htpb]
\centering
\subfigure[electrons]{\includegraphics[width=0.48\linewidth]{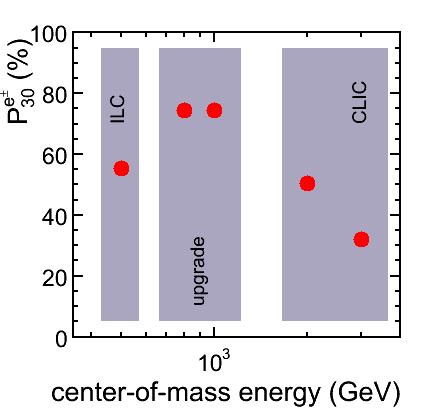}}
\subfigure[Higgs boson]{\includegraphics[width=0.48\linewidth]{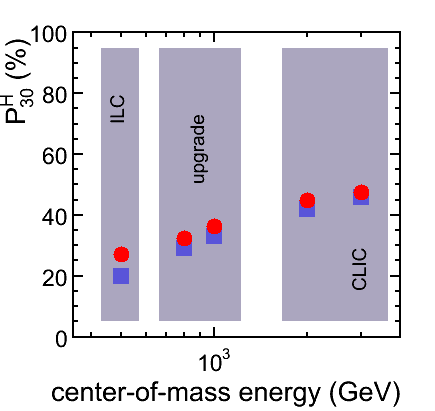}}
\caption{MadGraph~\cite{madgraph} prediction for the fraction of outgoing electrons and positrons with $ \theta < $ 30$^\circ$ in the Z-boson fusion process (leftmost figure). In the rightmost figure the fraction of Higgs bosons emitted in the forward direction is indicated for the W-boson fusion process (filled markers) and the Z-boson fusion process (open markers). }
\label{fig:p30_vs_sqrts_wwh_zzh}
\end{figure}

A precise reconstruction of the outgoing electrons in the Z-boson fusion process allows a measurement of the Higgs mass using the recoil scheme. Contrary to the Z-decay products in the Higgs-strahlung process, however, the electrons are very strongly peaked in the forward direction. At a {\em minimal} 500~GeV LC, 60 \% of electrons is emitted in the forward detector. Beyond 800~GeV the fraction of electrons that do not exceed a polar angle of 5$^\circ$ becomes quite significant, leading to a decrease of $P^{e^{\pm}}_{30} $ for even larger center-of-mass eenergy. The fraction of electrons (or positrons) in the forward detector is plotted as a function of center-of-mass energy in figure~\ref{fig:p30_vs_sqrts_wwh_zzh}. A more detailed discussion is found in reference~\cite{clicphysics}, where the requirement that this channel poses on the forward tracking system is clearly acknowledged.


The recoil mass reconstruction is not applicable to the $ e^+ e^- \rightarrow H \nu_e \bar{\nu}_e $ process of diagram (c). However, the large cross-section at large center-of-mass energy may turn out to be particularly relevant if a heavy Higgs boson with Standard Model-like couplings is found (clearly, some new physics mechanism must be invoked to avoid a contradiction with the fit to electroweak precision data~\cite{Z-pole}). In this process the polar angle distribution of the Higgs boson shows a clear preference for the forward region. The probability to find the Higgs boson in the forward detector $ P^{H}_{30} $, in the rightmost panel of figure~\ref{fig:p30_vs_sqrts_wwh_zzh}, ranges between approximately 30 \% at the ILC to 50 \% for CLIC. In case of two-body decay of the (light) Higgs boson, the polar angle of each of the decay products is strongly correlated with that of the Higgs boson. The multiple decay modes thus inherit the preference for the forward region. The identification of heavy flavour (bottom and charm tagging and $ \tau $-identification) leads to strong requirement on the vertex reconstruction performance of the detector.

The study of the properties and decay modes of the Higgs boson in a 500~GeV to 3~TeV is a rich field. The forward region is close to irrelevant for the most notorious benchmark analysis, the recoil mass measurement of a light SM Higgs boson produced through the Higgs-strahlung process. However, there are many other analyses and many other scenarios where the forward region may well gain considerable protagonism.

\section{Discussion}
\label{sec:conclusions}

The balance between forward and central detector performance is one of the most challenging aspects of the design of experiments for the next generation of $ e^+ e- $ colliders in the center-of-mass energy range from 0.5 to 3~TeV. In this note we have explored the polar angle distributions of a number of relevant physics processes. For each of these we have calculated the fraction of final state objects emitted in the forward detector. The very forward region ($ \theta < $ 5$^\circ$) is not considered here. 

The list of studied benchmark reactions is necessarily incomplete. In particular, we have not looked into some channels that are known to make a good case for forward tracking, like the forward-backward asymmetry for $ b \bar{b} $ and $ c \bar{c} $~\cite{demarteau_beijing,benchmarks}, the detection of scalar $ \tau $-leptons with a mass close to that of the neutralino~\cite{bambade} and the center-of-mass energy determination using $ e^+ e^- \rightarrow \mu^+ \mu^- \gamma $~\cite{benchmarks}. We nevertheless believe that the examples chosen here illustrate a quite general physics case. 

For the $ e^+ e^- \rightarrow Z/\gamma^*  \rightarrow l^+ l^- $ process at the Z-pole the final state products are predominantly emitted in the central detector. ISR at larger center-of-mass energies causes an increasing fraction of final state fermions to be emitted in the forward or backward direction.

With increasing center-of-mass energy, four-fermion, six-fermions and even eight-fermion processes will be abundantly produced. The example of $ t \bar{t} \rightarrow W^+ b W^- \bar{b} $ production clearly shows that such events are rarely contained fully in the central detector.

We have evaluated scalar electron production as an example of a process with a strong {\em t-channel} contribution. The polar angle distribution of the final state electrons quite generally shows pronounced peaks in the forward and backward direction.

We furthermore found the polar angle distribution of the final state fermions in di-boson production processes to be strongly peaked in the forward direction. The same is true for Higgs boson production through vector-boson fusion.

We have seen that in several processes diagrams exist with outgoing electrons that are not open to other fermions. In our examples (scalar electron production, Higgs production through Z-boson fusion) these diagrams lead to a significantly enhanced cross-section for final states with electrons. Even more important for this discussion is that the polar angle distribution of these diagrams has a very strong preference for the forward detector. The reconstruction of energetic electrons in the polar angle region from 5$^\circ $ to 30$^\circ $ therefore deserves special attention in the detector design.

While electrons make for a particularly strong argument, the design of the forward tracking region should not be limited to this one case. In several of our examples leptons and quarks of all flavours have a large probability to be emitted in the forward detector. In the examples of the $ t \bar{t} $ production and the $ H \nu_e \bar{\nu}_e $ process the identification of heavy flavour is particularly relevant. Thus, these channels may be used to derive requirements on the vertexing performance of the detector at small polar angle.

Finally, in (nearly) all our examples the relevance of the forward region increases with increasing center-of-mass energy of the collider. Therefore, as the next generation of $ e^+ e^- $ colliders will enter the the regime of 1~TeV or even several TeV, the focus of detector design may have to shift from the central detector to a more even balance between the central and forward detector systems.

\section{Acknowledgements}

The authors would like to acknowledge support from the SiLC collaboration. Part of the reported work was funded under MICINN projects CSD2007-0042 and FPA2008-03564-E/FPA. 

\begin{footnotesize}
\bibliographystyle{h-elsevier2}
\bibliography{forward}{}
\end{footnotesize}
\end{document}